**Title: Temperature-dependent Magnetic Transitions in CoCrPt-Ru-CoCrPt Synthetic Ferrimagnets**

**Authors:** Bradlee Beauchamp and Ernesto E. Marinero

School of Materials Engineering, Neil Armstrong Hall of Engineering, Purdue University, 701 W Stadium Ave, West Lafayette, IN 47907

Corresponding Author: <eemarinero@purdue.edu>

**Abstract:**

The magnetic orientations and switching fields of a CoCrPt-Ru-CoCrPt synthetic ferrimagnet with perpendicular magnetic anisotropy have been studied in the temperature range from 2 K to 300 K. It was found that two sets of magnetic transitions occur in the CoCrPt-Ru-CoCrPt ferrimagnet across this temperature range. The first set exhibits three magnetic transitions in the 50 K – 370 K range, whereas the second involves only two transitions in the 2 K and 50 K range. The observed magnetic hysteresis curves of the synthetic ferrimagnet are explained using the energy diagram technique framework pioneered by Koplak *et al.* [1] which accurately describes the competition between interlayer exchange coupling energy, Zeeman energy, and anisotropy energy in the system. In this work we expand the framework to include synthetic ferrimagnets (SFMs) comprising higher perpendicular magnetic anisotropy materials and large (4X) interlayer exchange coupling energies which are promising for the development of ultrafast (ps) magnetic switching free layers in MTJ structures. Furthermore, we apply the analysis to predict SFM magnetic hysteresis curves in a temperature regime that includes temperature extrema that a synthetic ferrimagnet would be expected to reliably operate at, were it to be utilized as a free layer in a memory or sensor spintronic device.

**Introduction:**

Synthetic ferrimagnet (SFM) trilayers consist of two antiparallel ferromagnetic (FM) films separated by a thin non-ferromagnetic metallic interlayer. For the case of identical FM layers, if the films are dissimilar in thickness, the SFM structure will exhibit a net magnetic moment (uncompensated ferrimagnet). The interlayer exchange coupling energy (IECE) of the SFM varies with the interlayer thickness in an oscillatory fashion [2] and it has been attributed to various physical processes that include dipolar magnetostatic interactions and Ruderman-Kittel-Kasuya-Yosida (RKKY) coupling. First observed by Grünberg et al. [3], films exhibiting antiferromagnetic coupling were utilized shortly thereafter in magnetic sensor devices based on the giant magnetoresistance (GMR) observed in Fe/Cr antiferromagnet structures [4][5]. More recently, SFMs and antiferromagnets have been utilized in magnetic tunnel junctions (MTJs) to provide exchange bias to the reference layer (typically a single, magnetically hard ferromagnet) such that its magnetization is "fixed" [6,7]. SFM trilayers or antiferromagnets have been considered to replace the recording or reference layer of the MTJ [8,9]. When utilized in an MTJ device the strength of coupling can determine if the SFM is acting as a reference or free layer. The coupling strength is derived from measuring the magnetic field required for overcoming the IECE which renders the magnetization orientation of the individual layers to be parallel. It is noted that free layer SFM structures incorporated into MTJ devices have demonstrated lower critical switching currents than single FM free layers without negatively affecting thermal stability [10]. Additionally, we have proposed that SFM free layers can exhibit ultrafast switching speeds down to the picosecond time regime [11].

Most MTJ devices utilize CoFeB as the FM electrode due to high tunneling magnetoresistance measured when used with MgO tunneling barriers [12]. However, the maximum thickness of CoFeB exhibiting perpendicular magnetic anisotropy (PMA) is limited to around 1.5 nm [13]. The magnetization of CoFeB is also relatively high, which increases the charge current needed for spin-transfer torque switching. CoCrPt is a material of interest for MTJ applications due to its low magnetization and its large anisotropy [14], resulting in lower switching currents, improved thermal stability, and the use of thicker FM layers with concomitant process control improvements. In addition, the SFM configuration circumvents the materials-restrictive low magnetic damping requirement for selection of the FM thin film for MTJ devices [11].

It is essential to tailor the IECE and switching properties of the SFM structure for use in memory devices. However, the magnetic properties of the SFM are temperature dependent, and memory devices could be expected to operate under extreme conditions within the range of 200 K to 370 K. In this paper the IECE of CoCrPt-Ru-CoCrPt trilayer structures has been investigated from 2 K to 300 K. It has been observed by Koplak *et al*. [1] that with decreasing temperature, the hysteresis loops of SFMs vary dramatically, and these authors developed a formalism to describe the observed changes in the hysteresis loop as a function of temperature in a CoFeB-Ta-CoFeB synthetic antiferromagnet. They employ an energy balance approach that includes the Zeeman energy, the IECE, and energy barriers for switching arising from the effective magnetic anisotropy energy. It was found that the two main parameters controlling the switching behavior with decreasing temperature is the ratio of the magnetic moments of the two constituent ferromagnetic layers as well as the energy barrier for switching of each film, which is temperature dependent. In this paper the energy diagram technique introduced by Koplak et al. is used to describe the magnetic transitions measured in a CoCrPt-Ru-CoCrPt SFM as a function of temperature. This is compared with their results on the CoFeB-Ta-CoFeB SFM structure. Predictions are also made for the magnetic transitions of the CoCrPt-Ru-CoCrPt SFM at 200 K to 370 K to exemplify the practical use of the energy diagram technique for assessing the robustness of a potential sensor or memory device employing a SFM read layer.

**Materials and Methods:**

All films were deposited without substrate heating or bias in a magnetron sputter system with a base pressure < $10^{-7}$ Torr. The films were grown on oxidized silicon (100) substrates, with dimensions ~ 5mmx5mm. These small coupons were obtained by manual dicing of larger (2" diameter) Si wafers. Thus, the actual dimensions of each sample exhibited small variations. Area measurements for each coupon were performed with optical microscopy with a reference scale. The error in surface area determination is ~8%. The thin film structure consisted of the following: Ta(5 nm)/Ru(10 nm)/CoCrPt(1.7 nm)/Ru(X nm)/CoCrPt(1.3 nm)/Ru(5 nm). The CoCrPt sputtering target has a nominal composition of $Co_{70}Cr_{18}Pt_{12}$. The Ta/Ru seed layer was used to promote CoCrPt growth with its basal (002) plane parallel to the thin film plane (c-axis out of plane). Magnetic hysteresis loops were collected using a Quantum Design MPMS-3 superconducting quantum interference device (SQUID) magnetometer with $10^{-8}$ emu sensitivity. All samples were mounted in the magnetometer using a plastic straw attachment. Measurements at 370 K could not be performed using this method due to errors on account of warping of the straw mount at this temperature. All magnetic hysteresis loops presented in this work were performed with the substrate aligned perpendicular to the direction of the applied magnetic field.

**Results and Discussion:**

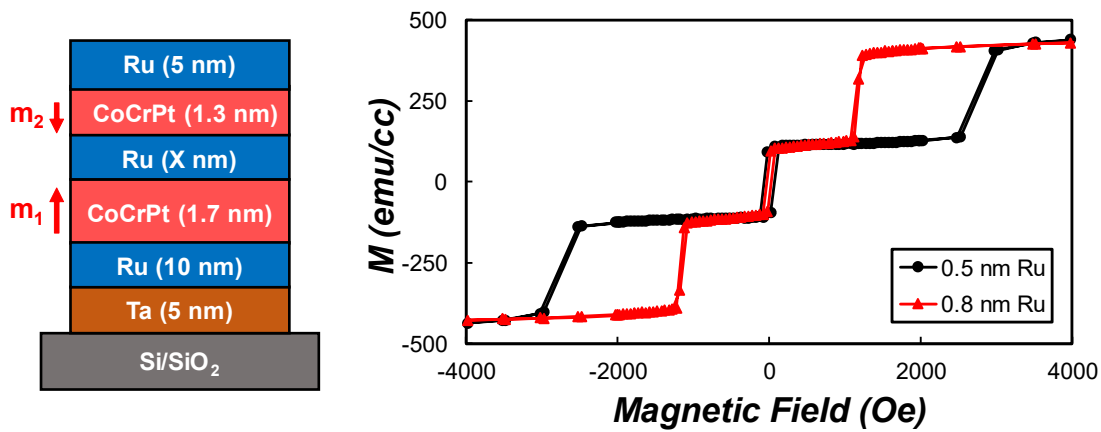

Figure 1. Left: Schematic representation of the film stack cross-section: red arrows indicate the direction of magnetization of the constituent FM layers at remanence. Right: Hysteresis curves of the SFM structures with two different Ru interlayer thicknesses measured at 300 K.

As shown in Fig. 1, the IECE of the CoCrPt-Ru-CoCrPt SFM can be tailored by varying the Ru interlayer thickness. The IECE per unit area of a SFM with dissimilar FM layers was estimated by Koplak et al. using the expression: $J_{EX} = -H_B m_2/S$. Here $H_B$ is the bias field which indicates the center of the outer loop. In Fig. 2a-b, $H_B$ is labeled and measured by finding the center of the outer loops at which point the SFM becomes saturated. At 2 K (Fig. 2c), there is no outer loop unlike in Fig. 2a-b. In this case a minor loop must be taken to locate the $H_B$ field, shown by the red curve in Fig. 2c. Peng et al. [15] performed similar minor loop measurement to determine the exchange coupling field in CoCrPt(18 nm)/Co(1 nm)/Ru(0.9 nm)/CoPtCr(8 nm) AFM structures as the magnetic measurements of these samples exhibited no outer loops present in the major loop of the hysteresis curve. A comparable measurement was later performed by Bandiera et al. [9], where a minor loop was measured to deduce the exchange coupling field in a SFM structure comprising Co/Pt multilayers, Co interlayers, Ru and CoFeB. Both studies demonstrate that the exchange coupling field can be obtained by the reversal of the softer magnetic layer in the SFM. Peng et al. measured this minor loop in the first quadrant of the hysteresis loop, while Bandiera et al. measured the minor loop in the third quadrant.

In the $J_{EX}$ equation, $m_2$ is the magnetic moment from the thinner magnetic layer ($m_1$ being the moment from the thicker layer), and S is the surface area of the film. The measured surface areas, S, are determined to be $0.21 \pm 0.01$ cm$^2$ and $0.09 \pm 0.01$ cm$^2$ for the hysteresis loops shown in Fig. 1 for the SFMs with 0.5 nm and 0.8 nm Ru spacer layers, respectively. The corresponding $J_{EX}$ values are calculated -0.07 erg/cm$^2$ and -0.03 erg/cm$^2$ for Ru=0.5 nm and 0.8 nm respectively. We note that the magnitude of IECE stronlgy depends not only on the Ru interlayer thickness, but also on the magnetic and structural properties of the Ru/CoCrPt interface. In particular, Peng et al. [15] in their study of CoCrPt(18 nm)/Ru(0.9 nm)/CoCrPt(8 nm) AFM system, demonstrated that the addition of a Co(1 nm) interlayer between Ru and CoPtCr, incremented $J_{EX}$ from 0.13 erg/cm$^2$ to 0.8 erg/cm$^2$. Generally, higher IECE is desirable for memory applications, including novel devices such as a double MTJ containing two SFM reference layers and a SFM free layer discussed in [11] which is predicted to switch in ps time scales.

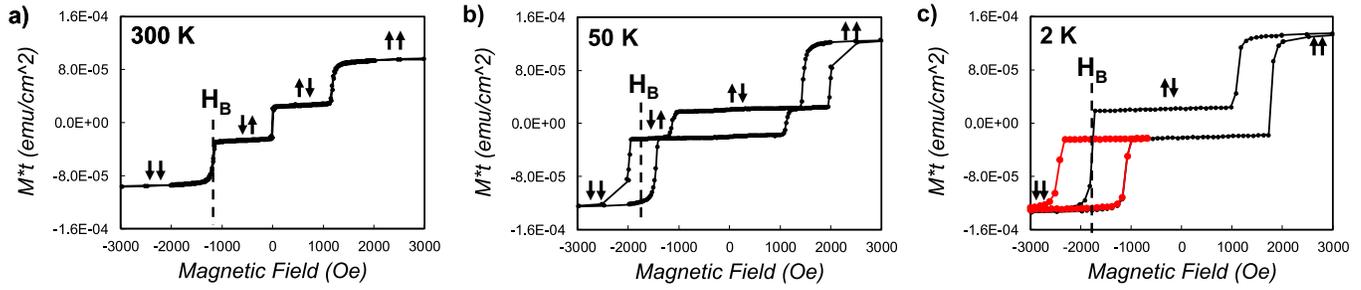

Figure 2. Hysteresis curves of CoCrPt(1.7)/Ru(0.8)/CoCrPt 1.3) SFM at a) 300 K, b) 50 K, and c) 2 K. The switching behavior from 50-300 K includes three magnetic transitions while two transitions are observed at 2 K. In Figs. 2a and 2b, the center of the outer loops is indicated by $H_B$. In Fig. 2c, the $H_B$ indicates the center of the minor loop (red curve) associated with the switching of the thinner magnetic layer.

An SFM has four possible magnetic configurations (↑↑, ↑↓, ↓↑, ↓↓), where the left and right arrows indicate the bottom (thicker) and top (thinner) FM layers, respectively as shown in Fig. 2. The number of transitions found in the hysteresis loop at a given temperature depend on the IECE, the Zeeman energy, and the energy barrier for magnetic reversal. From the literature, it is also evident that the magnetic field sweeping rate influences the magnetic switching behavior [16]. However, in this work each data point of the hysteresis curve is collected once the applied field has stabilized. Also, each hysteresis curve is collected once the sample temperature has fully stabilized.

Koplak et al. found for the CoFeB-Ta-CoFeB SFM three types of hysteresis loops over the temperature range studied. In the 180 K-300 K range, three magnetic transitions were observed for Type I hysteresis (↑↑-↑↓, ↑↓-↓↑, ↓↑-↓↓). Two transitions are present between 120 K-170 K for Type II hysteresis (↑↑-↑↓, ↑↓-↓↓) and 2 K-110 K (↑↑-↓↑, ↓↑-↓↓) for Type III hysteresis. These transitions are observed when the applied magnetic field is swept from positive to negative. The reverse transitions are encountered when the field is swept from negative to positive. In the case of the CoCrPt-Ru-CoCrPt system, we observe three transitions (↑↑-↑↓, ↑↓-↓↑, ↓↑-↓↓) in the 50 K to 300 K range (Fig. 2a-b). Whereas the number of magnetic transitions reduces to two when the sample is cooled down to 2 K (Fig. 2c), (↑↑-↑↓, ↑↓-↓↓) similar to the hysteresis loop measured in the 120 K-170 K range in the CoFeB-Ta-CoFeB system. The third set of magnetic transitions (↑↑-↓↑, ↓↑-↓↓) that are reported for the CoFeB-Ta-CoFeB SFM (Type III hysteresis) are not observed for the CoCrPt-Ru-CoCrPt SFM. To understand the differences in magnetic transitions between the SFM system here reported (CoCrPt-Ru-CoCrPt) and that studied by Koplak et al. (CoFeB-Ta-CoFeB), we provide a comparative analysis based on the magnetic properties of the constituent layers and on the higher IECE (4X at 300 K) provided by the Ru in our study. An important contribution of this study is the extension of the energy balance formulation developed by Koplak et al. to SFM structures comprising higher perpendicular magnetic anisotropy ferromagnets that are critical for thermally stable single nm scale magnetic structures. Table 1 summarizes the types of hysteresis curves observed in both the CoCrPt-Ru-CoCrPt and the CoFeB-Ta-CoFeB SFMs as a function of temperature.

| Hysteresis Type and Magnetic Transitions | CoCrPt-Ru-CoCrPt SFM | CoFeB-Ta-CoFeB SFM |
|---|---|---|
| Type I Hysteresis ↑↑-↑↓, ↑↓-↓↑, ↓↑-↓↓ | 50 - 300 K | 180 - 300 K |
| Type II Hysteresis ↑↑-↑↓, ↑↓-↓↓ | 2 - 40 K | 120 - 170 K |
| Type III Hysteresis ↑↑-↓↑, ↓↑-↓↓ | - | 2 - 110 K |

In Table 1, the hysteresis types and the associated magnetic transitions observed in the CoCrPt-Ru-CoCrPt SFM are shown in the second column and they are compared to the CoFeB-Ta-CoFeB SFM results reported by Koplak *et al.* [1] in the third column. The indicated magnetic transitions occur when the magnetic field is swept from positive saturation to negative saturation. The bold arrows represent the magnetic moment, $m_1$, of the thicker magnetic layer. The temperature range where each type of hysteresis curve is observed are provided under the heading of the different SFM structures. Three magnetic transitions are reported by Koplak et al. for the CoFeB-Ta-CoFeB, while two magnetic transitions are observed for the CoCrPt-Ru-CoCrPt SFM.

Here we analyze the magnetic transitions exhibited by the SFM with a 0.8 nm Ru spacer (Fig. 1) using the energy diagram technique. This technique relies on a simple energy balance (Eq. 1) which contains the IECE ($E_{EX}$), Zeeman energy ($E_Z$), and the potential barriers $E_{eff1}$ and $E_{eff2}$ (corresponding to the 1.7 nm and 1.3 nm CoCrPt, respectively).

**Eq. 1)** $E_{Total} = E_{EX} + E_Z + E_{eff1} + E_{eff2}$

The IECE, $E_{EX}$, is proportional to the surface area of the sample and can be estimated from $|E_{EX}| = H_B \cdot m_2$. Here $H_B$ represents the bias field, which is measured by locating the center of the minor loop of the softer magnet as previously described. In the case of the first type of switching shown in Figs. 3a and 3b, there are three loops: the center field of the outer loops is $H_B$ and indicates the strength of IECE. The potential energy barrier separates the two perpendicular orientations of magnetization. Notably the hysteresis curves are measured along the easy-axis, therefore, $E_{eff1}$ and $E_{eff2}$ are not equal to the anisotropy energy determined from the hard axis hysteresis. In a hysteresis loop with three transitions (Figs. 3a and 3b), $E_{eff1}$ can be

estimated from the coercive field of the outer loops as $H_{C-outer} = \frac{E_{eff1}}{2 \cdot m_1}$. Then $E_{eff2}$ can be calculated from the coercive field of the inner loop expressed by $H_{C-inner} = \frac{E_{eff1}+E_{eff2}}{2(m_1-m_2)}$. These estimates arise from the equations derived by Koplak et al. describing the possible magnetic transitions. The Zeeman energy, $E_Z$, is proportional to the applied magnetic field and can be expressed as $E_Z = -(m_1 + m_2) \cdot H$.

The second type of hysteresis curve shown in Fig. 3c has no outer loops, which are needed to estimate $E_{EX}$. However, one can still calculate $H_B$ and thus $E_{EX}$ by measuring the minor loop as shown in Fig. 2c. This is obtained by switching the softer, thinner $m_2$ magnetic layer after the SFM has been saturated [9,15]. This method of measuring the bias field advances the application of the energy balance framework to SFM structures where there are no outer loops in the hysteresis curve. $H_B$, labeled in Fig. 2c, was determined to be -1798 Oe. The minor loop in Fig. 2c is measured by saturating the SFM to the ↓↓ orientation, sweeping the magnetic field to just beyond the ↓↓-↓↑ magnetic transition, and then saturating the SFM back to the ↓↓ orientation. This indicates an IECE of -0.11 erg/cm² for the SFM at 2 K. After $E_{EX}$ is obtained, $E_{eff1}$ and $E_{eff2}$ can be calculated using equations $H_{\uparrow\uparrow-\uparrow\downarrow} = \frac{2|E_{EX}|-E_{eff2}}{2m_2}$ and $H_{\uparrow\downarrow-\downarrow\downarrow} = -\frac{2|E_{EX}|+E_{eff1}}{2m_1}$, respectively. The resulting energy diagram (Fig. 3c) is consistent with the transition fields of the minor loop and the saturated hysteresis loop. $E_{eff1}$ and $E_{eff2}$ are plotted at each temperature in Fig. 5a. We note that, $E_{eff1}$ is larger than $E_{eff2}$ until the temperature is lowered to 2 K, most likely due to the changing $m_1/m_2$ ratio as temperature is decreased. It is noted that the error associated with the $E_{eff}$ parameters is around ± 1x10⁻⁵ erg so the observed difference in their values is significant. A similar behavior was observed by Koplak et al. which is observed at 100 K and is attributed to the changing ratio of $m_1$ to $m_2$ as temperature is decreased.

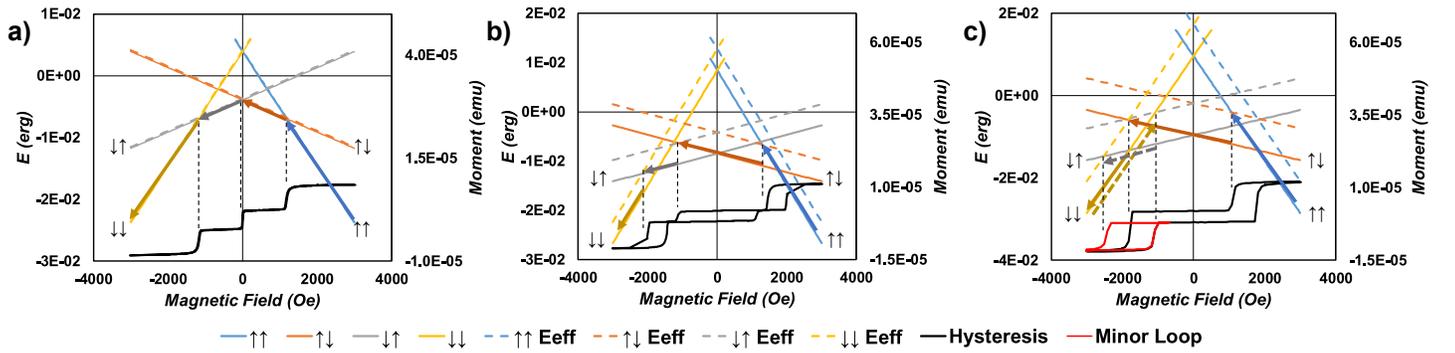

Figure 3. Energy diagrams of the CoCrPt(1.7)/Ru(0.8)/CoCrPt(1.3) SFM at a) 300 K, b) 50 K, and c) 2 K. The solid lines indicate the total energy, excluding the energy barriers, while the dashed lines include the temperature-dependent energy barrier term, $E_{eff}$. The hysteresis curves are shown in each pane with corresponding

magnetic moments on the secondary axis. The red hysteresis loop in c) displays the minor loop measured to determine $H_B$. Dashed arrows indicate the energies associated with the minor loop and the corresponding transitions.

The energy diagrams shown in Fig. 3 describe the magnetic transitions occurring for each temperature. Solid lines indicate the energy of the SFM system with zero $E_{eff}$, i.e. when there are no energy barriers to overcome. The solid lines are thus, the addition of the Zeeman energy and the IECE, with a y-intercept equal to the IECE. The dashed lines represent the total energy of the system after the $E_{eff1}$ and $E_{eff2}$ are included, as described by Eq. 1. As the magnetic field is swept, the magnetic orientation present is the one with the lowest energy. Shown in Fig. 3a, the blue solid line represents the ↑↑ orientation as the field is lowered from +2 T. If the energy barrier for reversal of each layer is zero, the ↑↑-↑↓ will occur at the intersection of the blue solid line and the orange solid line representing the total energy of the ↑↓ orientation. At 300 K, the $E_{eff}$ energies for $m_1$ and $m_2$ are negligibly low such that the magnetic transitions occur approximately at the solid line intersections.

The potential barriers become larger as the temperature is decreased to 2 K (Fig. 5a). Since the $E_{eff}$ terms are not field-dependent, they shift the dashed lines up along the y-axis. For a magnetic transition to occur, the magnetic field must be changed such that the energy barrier between the solid and dashed line is crossed. As seen in Fig. 3b, the ↑↑-↑↓ transition no longer occurs at the intersection of the solid blue and orange lines, but at the point where the potential barrier of another orientation energy is crossed. As the potential barriers increase with lower temperature, certain transitions are prohibited from occurring due to the existence of lower energy states from other magnetic orientations. At 2 K (Fig. 3c), the ↑↓-↓↑ transition does not occur as it does at 300 K and 50 K (Fig. 3a-b) since the potential barrier of the ↓↓ state is lower in energy than the ↓↑ state.

The third set of magnetic transitions (↑↑-↓↑, ↓↑-↓↓), or Type III hysteresis, occurs when the condition $E_{eff1} < E_{eff2} \cdot \frac{m_1}{m_2} - 2|E_{EX}| \cdot \frac{m_1-m_2}{m_2}$ is satisfied [1]. The CoCrPt-Ru-CoCrPt SFM studied does not meet this requirement and does not show this set of transitions even down to 2 K. Compared to the CoFeB-Ta-CoFeB SFM reported by Koplak *et al.* [1], which has an IECE at 300 K of $E_{EX}/S$ = -0.01 erg/cm$^2$, the CoCrPt-Ru-CoCrPt SFM has an IECE at 300 K of $E_{EX}/S$ = -0.04 erg/cm$^2$. The CoFeB-Ta-CoFeB SFM and the CoCrPt-Ru-CoCrPt SFM have $m_1/m_2$ ratios of 1.38 and 1.79, respectively. At 300 K the effective anisotropy energy barriers for both systems are: $E_{eff1}/S$ = 4.0x10$^{-3}$ erg/cm$^2$ and $E_{eff2}/S$ = 2.5x10$^{-3}$ erg/cm$^2$ for CoFeB-Ta-CoFeB, $E_{eff1}/S$ = 1.7x10$^{-3}$ erg/cm$^2$ and $E_{eff2}/S$ = 0.73x10$^{-3}$ erg/cm$^2$ for the CoCrPt-Ru-CoCrPt. The energy barriers for the CoCrPt-Ru-CoCrPt are lower for $E_{eff1}$ and $E_{eff2}$ by a factor of 2.3 and 3.4, respectively. This disparity in $E_{eff}$ causes the right side of the inequality to be lower, which explains why the third set of magnetic transitions are absent in the CoCrPt-Ru-CoCrPt SFM.

Figure 4 illustrates the dependence of the left and right sides of the inequality $E_{eff1} < E_{eff2} \cdot \frac{m_1}{m_2} - 2|E_{EX}| \cdot \frac{m_1-m_2}{m_2}$ on the IECE for both the CoCrPt and CoFeB SFM systems. The plot shows the energies calculated at 100 K, since CoFeB-Ta-CoFeB shows the third set of magnetic transitions at this temperature. At a given IECE, the third set of magnetic transitions should be observed when the right side of the equation is larger than $E_{eff1}$. When plotted as a function of $E_{EX}$, the right side of the inequality is a line whose slope is determined by the $m_1/m_2$ ratio and the intercept by the product of $E_{eff2}$ and $m_1/m_2$. The dependence of the right side of the inequality on $m_1/m_2$ is also illustrated in Fig. 4a-b. As $m_1/m_2$ approaches 1, the slope and the y-axis intercept of the right side of the inequality is lowered. Since it has been shown that fast spin transfer torque switching can be achieved with a low $m_1/m_2$ ratio [11], this analysis is important in understanding the type of hysteresis curves that will be present in the SFM when tailoring the ratio of magnetic moments. It is evident from Fig. 4a that the magnetic switching behavior of the CoCrPt-Ru-CoCrPt SFM will not exhibit the third type of magnetic switching for either stronger or weaker IECE as the intercept of the right side of the inequality is lower than $E_{eff1}$.

As mentioned earlier, the SFM can be used as a replacement for a single FM layer in a memory device. Such devices are expected to operate successfully over a wide range of temperatures. Therefore, it is important to predict the behavior of the SFM at any temperature. The energy diagram technique can be used to predict the transition fields of the SFM if the temperature dependence of $E_{eff}$, $E_{EX}$, and the magnetization, m, are known. Figure 5a-c shows the temperature dependence of $E_{eff}$, $E_{EX}$, and m, respectively. $E_{eff}$, $E_{eff1}$, and $E_{eff2}$ are proportional to $m^{\frac{n(n+1)}{2}}$ at lower temperatures (<150 K), while at higher temperatures the potential energy barriers are proportional to $m^n$ similar to the temperature dependence of the magnetic anisotropy observed for other materials [17,18]. Here, m is the magnetic moment and n is the exponent of the magnetic anisotropy function (n=2 is typical for uniaxial anisotropy). Fig. 5a shows the fit for $E_{eff2}$ based on the $m^{\frac{n(n+1)}{2}}$ proportionality. Good agreement is observed with the fit until around 150 K, above which there are significant differences between the measured $E_{eff}$ and the fit. At higher temperatures, $E_{eff2}$ is proportional to $m^n$ as predicted in [17]. $E_{eff1}$ shows a similar temperature dependence as $E_{eff2}$ and the same exponent for the magnetic anisotropy function fits the data, however the data point at 50 K significantly deviates from the trend. The parameters $E_{EX}$ and m change linearly in this temperature range. The m values are later extrapolated to 370 K using this linear fit. This is appropriate since this temperature is well below the Curie temperature reported for $Co_{70}Cr_{18}Pt_{12}$ (~ 673 K [19]), around which the temperature dependence of m would significantly depart from the linear trend. It is noted in Fig. 5c that the $m_1/m_2$ ratio across the temperature range studied is larger than the value of 1.3 expected from the nominal layer thickness of 1.7 nm and 1.3 nm. As an example, at 2 K and 300 K the ratio is 1.38 and 1.7, respectively. The discrepancy may result from the manner in which the magnetic moments are determined for the results presented in Fig. 5c, namely, they are estimated from the magnetization measurements of the SFM hysteresis loops at saturation and remanence. It is feasible that at remanence, contributions from the thin film multi-domain magnetic state may influence the magnitude of the measured magnetization. In addition, whereas the film thicknesses and moments of the constituent layers were determined in samples grown on Ta(5 nm)/Ru(10 nm) underlayers, in the SFM, $m_2$ is grown on thin Ru, therefore its crystalline quality may differ from $m_1$. This could also result in differences in magnetic properties of the layers as a function of temperature.

Fitting the trends seen in Fig. 5 allows one to predict the behavior at 200 K and 370 K, the temperature extrema that a spintronic sensor or memory device could potentially expected to operate reliably. The energy diagrams for the CoCrPt(1.7)/Ru(0.8)/CoCrPt(1.3) SFM at these two temperatures are shown in Fig. 6. The energy diagram in Fig. 6a is interpolated from the parameters depicted in Fig. 5, while the energy diagram in Fig. 5b is extrapolated from fitted parameters in Fig 5. These energy diagrams are constructed using the fitted parameters from Fig. 5. Both energy diagrams in Fig. 6a and 6b show magnetic transitions corresponding to the type I regime [1]. The transition field for ↑↑-↑↓ (where $m_1$ reversal occurs) is predicted to be 1400 Oe and 950 Oe at 200 K and 370 K, respectively. In Fig. 6a the hysteresis diagram measured at 200 K is shown and agrees with the predicted transitions from the energy diagram. The hysteresis curve at 370 K was not be acquired due to errors introduced by deformation of the straw sample mounting arrangement employed for all other measurements as discussed in the materials and methods section.

The CoCrPt-Ru-CoCrPt SFM system provides significant advantages over CoFeB-Ta-CoFeB as free layers in MTJ structures. These derive from the lower saturation magnetization and its superior magnetic anisotropy [14]. The anisotropy in CoCrPt is largely determined by its magnetocrystalline anisotropy as opposed to interface anisotropy for the case of CoFeB. Therefore, CoCrPt films exhibit PMA for film thicknesses up to 15 nm. Thus, the $m_1/m_2$ ratio in the films can be controlled more precisely, allowing for more tunability of the SFM properties. This is of particular interest for the implementation of ps magnetic switching employing SFM structures as proposed by Camsari et al. [11]. Said SFMs require larger IECE than that provided by Ta layers in order to achieve ps magnetic switching. Therefore, it is important to extend the energy diagram technique developed by Koplak et al. to this material structure to understand its behavior at temperature regimes of interest for practical utilization of spintronic memory devices exploiting these SFMs.

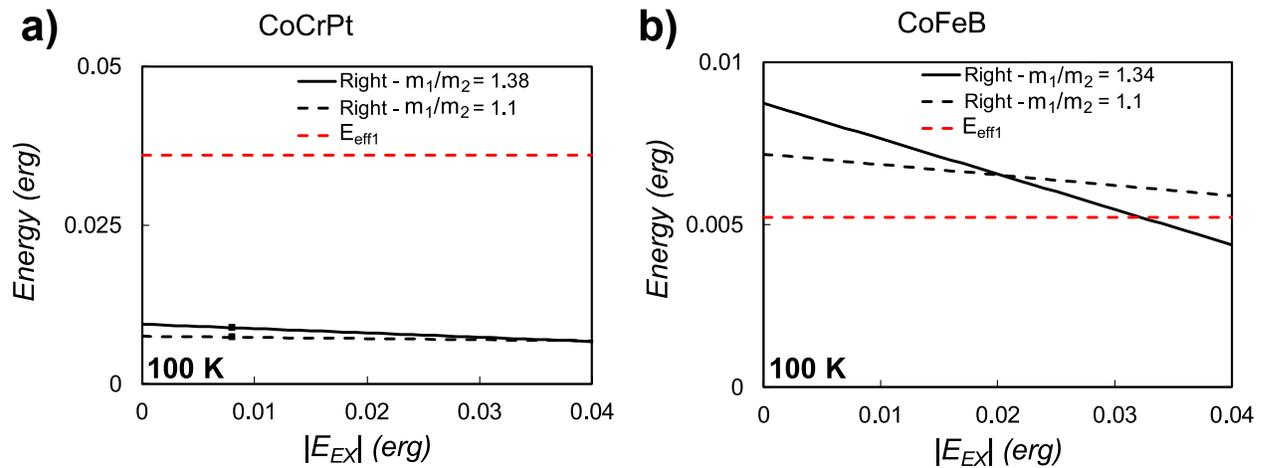

Figure 4. The left and right sides of the inequality $E_{eff1} < E_{eff2} \cdot \frac{m_1}{m_2} - 2|E_{EX}| \cdot \frac{m_1-m_2}{m_2}$ plotted as a function of $E_{EX}$ for the a) CoCrPt-Ru-CoCrPt and b) CoFeB-Ta-CoFeB SFM. The black solid and dashed lines labeled "Right – $m_1/m_2$" represents the right side of the inequality at different magnetic ratios. The left side of the inequality is represented by the red curve and is labeled $E_{eff1}$. The black squares shown in a) represent the observed IECE of the CoCrPt-Ru-CoCrPt SFM at 100 K.

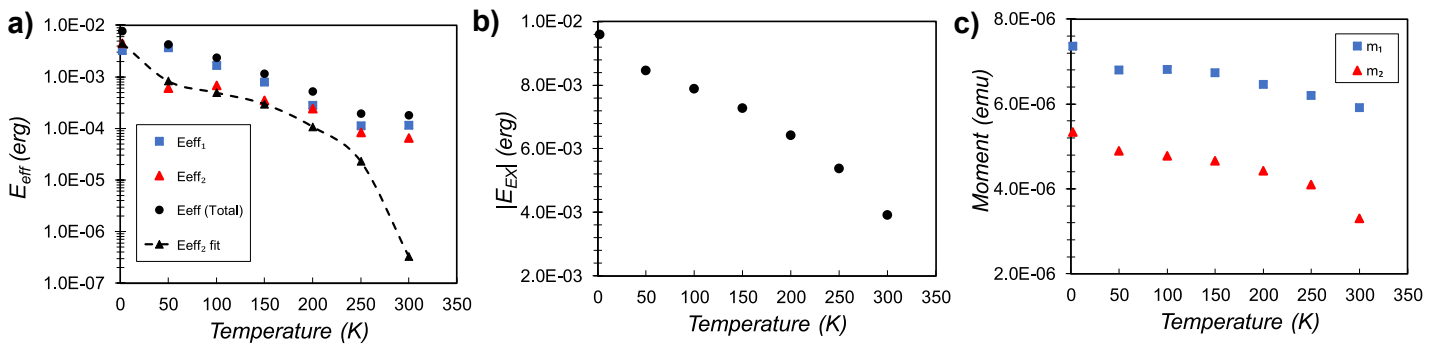

Figure 5. a) $E_{eff1}$, $E_{eff2}$, and $E_{eff}$ (Total) plotted versus temperature. b) $|E_{EX}|$ plotted versus temperature. c) The magnetic moments, $m_1$ and $m_2$ of the 1.7 nm and 1.3 nm thick CoCrPt layers, respectively, plotted versus temperature.

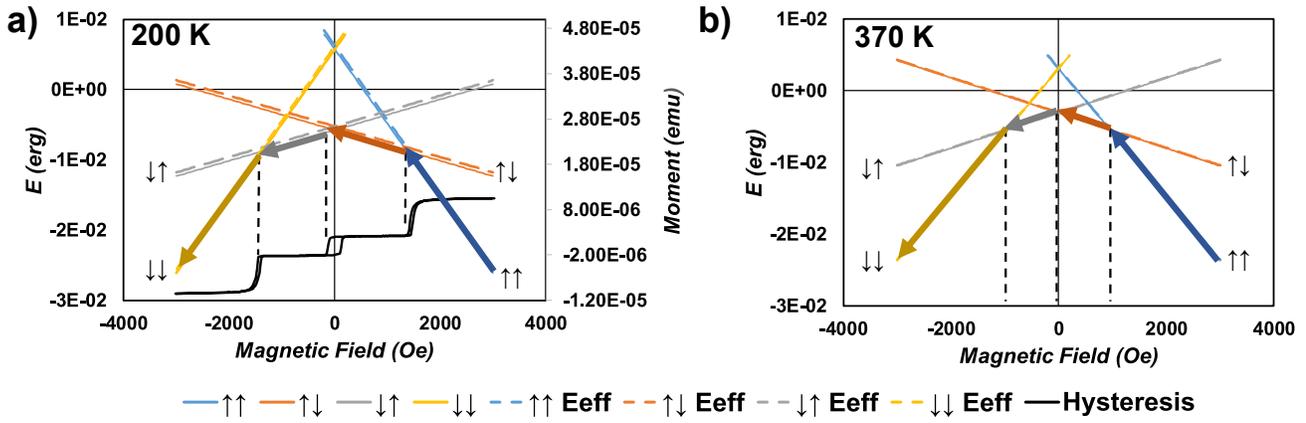

Figure 6. The energy diagram for the CoCrPt(1.7)/Ru(0.8)/CoCrPt(1.3) SFM at 200 K and 370 K. The energy diagram in a) is interpolated from the parameters derived in Fig. 5, while the energy diagram in b) is extrapolated from fitted parameters in Fig 5. The transition fields are indicated by the vertical dashed lines. The hysteresis curve measured at 200 K is shown in a).

**Conclusions:**

MTJ devices incorporating SFM structures as free layers are expected operate reliably over a wide temperature range that under extreme conditions could range from 200 K to 370 K. Therefore, it is important to determine the temperature dependence of their magnetic transitions to engineer key material properties such as IECE and the energy barriers associated with magnetization reversal. In this paper the IECE of CoCrPt/Ru/CoCrPt SFMs has been investigated from 2 K to 300 K. Building on previous work by Koplak et al. [1], this work further elucidates the temperature-dependence of the potential barrier for magnetization reversal. The magnitude of $H_B$ was derived from minor loop measurements, as the hysteresis curves exhibited no outer loops, a procedure reported also in refs. [9,15]. $H_B$ is needed to estimate IECE and details of the energy diagram technique first described by Koplak et al. [1]. In this stduy, we observe two types of hysteresis curves in CoCrPt/Ru/CoCrPt SFMs: one above 50 K with three subloops and the other at 2 K with two magnetic transitions. Type III hysteresis, seen in the CoFeB-Ta-CoFeB SFM system in Koplak et al., is not observed in the CoCrPt/Ru/CoCrPt SFM, due to the large $E_{eff1}$ which prevents the $E_{eff1} < E_{eff2} \cdot \frac{m_1}{m_2} - 2|E_{EX}| \cdot \frac{m_1 - m_2}{m_2}$ inequality from being satisfied. $E_{eff1}$ corresponds to the energy barrier between magnetization directions present in the CoCrPt film.

The utilization of the energy diagram technique to understand magnetic transitions in CoCrPt/Ru/CoCrPt SFMs enabled us to predict the magnetic orientations present in this SFM system at any given temperature. Future work will include studies in SFM structures comprising different ferromagnetic materials, alternative exchange coupling interlayers such as Rh to increment the magnitude of IECE, as well as the incorporation of nanoscale interlayer magnetic materials between the FM constituent layers and the non-magnetic spacer to enhance IECE, as demonstrated by Peng et al [15] will also be investigated. In addition, we plan to extend the temperature range investigated in this study and to utilize synchrotron structural and magnetic characterization probes to better characterize their structural and magnetic properties.

**Funding:**

This research was funded in part by Dr. Marinero's startup funds from Purdue University School of Materials Engineering and in part by the Office of Naval Research (Award# N00014-18-1-2481).

**References:**

[1] O. Koplak, A. Talantsev, Y. Lu, A. Hamadeh, P. Pirro, T. Hauet, R. Morgunov, S. Mangin, Magnetization switching diagram of a perpendicular synthetic ferrimagnet CoFeB/Ta/CoFeB bilayer, J. Magn. Magn. Mater. 433 (2017) 91–97. https://doi.org/10.1016/j.jmmm.2017.02.047.

[2] S.S.P. Parkin, N. More, K.P. Roche, Oscillations in Exchange Coupling and Magnetoresistance in Metallic Superlattices Structures: Cu/Ru, Co/Cr and Fe/Cr, Phys. Rev. Lett. 64 (1990) 2304–2308. https://doi.org/10.1103/PhysRevLett.64.2304.

[3] P. Grünberg, R. Schreiber, Y. Pang, M.B. Brodsky, H. Sowers, Layered Magnetic Structures: Evidence for Antiferromagnetic Coupling of Fe Layers across Cr Interlayers, Phys. Rev. Lett. 57 (1986) 2442–2445. https://doi.org/10.1103/PhysRevLett.57.2442.

[4] M.N. Baibich, J.M. Broto, A. Fert, F.N. Van Dau, F. Petroff, P. Etienne, G. Creuzet, A. Friederich, J. Chazelas, Giant Magnetoresistance of (001)Fe/(001)Cr Magnetic Superlattices, Phys. Rev. Lett. 61 (1988) 2472–2475. https://doi.org/10.1103/PhysRevLett.61.2472.

[5] G. Binasch, P. Grünberg, F. Saurenbach, W. Zinn, Enhanced magnetoresistance in layered magnetic structures with antiferromagnetic interlayer exchange, Phys. Rev. B. 39 (1989) 4828–4830. https://doi.org/10.1103/PhysRevB.39.4828.

[6] S. Parkin, X. Jiang, C. Kaiser, A. Panchula, K. Roche, M. Samant, Magnetically engineered spintronic sensors and memory, Proc. IEEE. 91 (2003) 661–679. https://doi.org/10.1109/JPROC.2003.811807.

[7] S.S.P. Parkin, C. Kaiser, A. Panchula, P.M. Rice, B. Hughes, M. Samant, S.H. Yang, Giant tunnelling magnetoresistance at room temperature with MgO (100) tunnel barriers, Nat. Mater. 3 (2004) 862–867.

https://doi.org/10.1038/nmat1256.

[8] D. Watanabe, Et Al., Interlayer exchange coupling in perpendicularly magnetized synthetic ferrimagnet structure using CoCrPt and CoFeB, J. Phys. Conf. Ser. 200 (2010) 72104. https://doi.org/10.1088/1742-6596/200/7/072104.

[9] S. Bandiera, R.C. Sousa, Y. Dahmane, C. Ducruet, C. Portemont, V. Baltz, S. Auffret, I.L. Prejbeanu, B. Dieny, Comparison of synthetic antiferromagnets and hard ferromagnets as reference layer in magnetic tunnel junctions with perpendicular magnetic anisotropy, IEEE Magn. Lett. 1 (2010). https://doi.org/10.1109/LMAG.2010.2052238.

[10] J. Hayakawa, S. Ikeda, K. Miura, M. Yamanouchi, Y.M. Lee, R. Sasaki, M. Ichimura, K. Ito, T. Kawahara, R. Takemura, T. Meguro, F. Matsukura, H. Takahashi, H. Matsuoka, H. Ohno, Current-induced magnetization switching in MgO barrier magnetic tunnel junctions with CoFeB-based synthetic ferrimagnetic free layers, in: IEEE Trans. Magn., 2008: pp. 1962–1967. https://doi.org/10.1109/TMAG.2008.924545.

[11] K.Y. Camsari, A.Z. Pervaiz, R. Faria, E.E. Marinero, S. Datta, Ultrafast Spin-Transfer-Torque Switching of Synthetic Ferrimagnets, IEEE Magn. Lett. 7 (2016) 1–5. https://doi.org/10.1109/LMAG.2016.2610942.

[12] S. Ikeda, J. Hayakawa, Y. Ashizawa, Y.M. Lee, K. Miura, H. Hasegawa, M. Tsunoda, F. Matsukura, H. Ohno, Tunnel magnetoresistance of 604% at 300 K by suppression of Ta diffusion in CoFeBMgOCoFeB pseudo-spin-valves annealed at high temperature, Appl. Phys. Lett. 93 (2008). https://doi.org/10.1063/1.2976435.

[13] S. Ikeda, K. Miura, H. Yamamoto, K. Mizunuma, H.D. Gan, M. Endo, S. Kanai, J. Hayakawa, F. Matsukura, H. Ohno, A perpendicular-anisotropy CoFeB – MgO magnetic tunnel junction, Nat. Mater. 9 (2010) 721–724. https://doi.org/10.1038/nmat2804.

[14] D. Watanabe, S. Mizukami, M. Oogane, H. Naganuma, Y. Ando, T. Miyazaki, Fabrication of MgO-based magnetic tunnel junctions with CoCrPt perpendicularly magnetized electrodes, J. Appl. Phys. 105 (2009) 126–129. https://doi.org/10.1063/1.3062816.

[15] Y. Peng, J.G. Zhu, D.E. Laughlin, Interfacial Co nanolayers for enhancing interlayer exchange coupling in antiferromagnetic interlayer exchange coupling media, J. Appl. Phys. 91 (2002) 7676–7678. https://doi.org/10.1063/1.1452262.

[16] R.B. Morgunov, E.I. Kunitsyna, A.D. Talantsev, O. V. Koplak, T. Fache, Y. Lu, S. Mangin, Influence of the

magnetic field sweeping rate on magnetic transitions in synthetic ferrimagnets with perpendicular anisotropy, Appl. Phys. Lett. 114 (2019). https://doi.org/10.1063/1.5096951.

[17] J.B. Staunton, L. Szunyogh, A. Buruzs, B.L. Gyorffy, S. Ostanin, L. Udvardi, Temperature dependence of magnetic anisotropy: An ab initio approach, Phys. Rev. B - Condens. Matter Mater. Phys. 74 (2006) 144411. https://doi.org/10.1103/PhysRevB.74.144411.

[18] B.D. Cullity, C.D. Graham, Introduction to Magnetic Materials, John Wiley & Sons, Inc., Hoboken, NJ, USA, 2008. https://doi.org/10.1002/9780470386323.

[19] M.F. Doerner, T. Yogi, D.S. Parker, S. Lambert, B. Hermsmeier, O.C. Allegranza, T. Nguyen, Composition Effects in High Density CoPtCr Media, IEEE Trans. Magn. 29 (1993) 3667–3669. https://doi.org/10.1109/20.281263.